\documentclass[11pt]{article}
\usepackage{amsmath,amssymb}

\def\xxx#1           {{\sf hep-th/#1} }

\def\npb#1(#2)#3     {Nucl. Phys. {\bf B#1} (#2) #3 }
\def\rep#1(#2)#3     {Phys. Rept. {\bf #1} (#2) #3 }
\def\pla#1(#2)#3     {Phys. Lett. {\bf A#1} (#2) #3 }
\def\plb#1(#2)#3     {Phys. Lett. {\bf B#1} (#2) #3 }
\def\prl#1(#2)#3     {Phys. Rev. Lett.{\bf #1} (#2) #3 }
\def\prd#1(#2)#3     {Phys. Rev. {\bf D#1} (#2) #3 }
\def\ap#1(#2)#3      {Ann. Phys. {\bf #1} (#2) #3 }
\def\rmp#1(#2)#3     {Rev. Mod. Phys. {\bf #1} (#2) #3 }
\def\cmp#1(#2)#3     {Comm. Math. Phys. {\bf #1} (#2) #3 }
\def\mpla#1(#2)#3    {Mod. Phys. Lett. {\bf A#1} (#2) #3 }
\def\ijmp#1(#2)#3    {Int. J. Mod. Phys. {\bf A#1} (#2) #3 }
\def\cqg#1(#2)#3     {Class. Quant. Grav. {\bf #1} (#2) #3 }
\def\am#1(#2)#3      {Adv. Math. {\bf #1} (#2) #3 }
\def\im#1(#2)#3      {Invent. Math. {\bf #1} (#2) #3 }
\def\jhep#1(#2)#3    {JHEP {\bf #1} (#2) #3 }
\def\npps#1(#2)#3    {Nucl. Phys. Proc. Suppl. {\bf #1} (#2) #3 }
\def\jgp#1(#2)#3     {J. Geom. Phys. {\bf #1} (#2) #3 }
\def\atmp#1(#2)#3    {Adv. Theor. Math. Phys. {\bf #1} (#2) #3}
\def\lmp#1(#2)#3     {Lett. Math. Phys. {\bf #1} (#2) #3}

\begin{document}
\thispagestyle{empty}
\null\vskip-24pt \hfill LAPTH-861/01 \vskip-10pt \hfill CERN-TH/2001-178

\begin{center}
\vskip 0.2truecm {\Large\bf
(2,0) Superconformal OPEs in D=6,\\Selection Rules and Non-renormalization Theorems}\\
\vskip 0.5truecm
B. Eden$^{*}$\footnote{email:{\tt burkhard@lapp.in2p3.fr}},
S. Ferrara$^\dagger$\footnote{email:{\tt sergio.ferrara@cern.ch}} and
E. Sokatchev$^{*}$\footnote{email:{\tt sokatche@lapp.in2p3.fr}}\\
\vskip 0.4truecm
{\it $^{*}$ Laboratoire d'Annecy-le-Vieux de Physique
Th{\'e}orique\footnote{UMR 5108 associ{\'e}e {\`a}
l'Universit{\'e} de Savoie} LAPTH, \\ Chemin de Bellevue, B.P. 110,
F-74941 Annecy-le-Vieux, France} \\
\vskip 0.3cm
{\it $^\dagger$ CERN, Theory Division, CH 1211 Geneva 23, Switzerland, and \\
Laboratori Nazionali di Frascati, INFN, Italy, and \\
Department of Physics and Astronomy, University of California, Los Angeles, CA 
90095, USA}
\end{center}

\vskip 1truecm \Large
\centerline{\bf Abstract} \normalsize We analyse the OPE of any
two 1/2 BPS operators of (2,0) SCFT$_6$ by constructing all
possible three-point functions that they can form with another, in
general long operator. Such three-point
functions are uniquely determined by superconformal symmetry. \\
Selection rules are derived, which allow us to infer
``non-renormalization theorems'' for an abstract superconformal
field theory. The latter is supposedly related to the
strong-coupling dynamics of $N_c$ coincident M5 branes, dual, in
the large-$N_c$ limit, to the bulk M-theory compactified on
AdS$_7 \times$S$_4$. \\
An interpretation of extremal and next-to-extremal correlators in terms of
exchange of operators with protected conformal dimension is given.

\newpage
\setcounter{page}{1}\setcounter{footnote}{0}

\section{Introduction}\label{intro}
In recent time, many tests (for reviews of the different
approaches see, e.g., \cite{Frrev,Brev,HWrev}) of the AdS/CFT
duality (see, e.g., \cite{AGMOO}) between bulk supergravity on
AdS$_{p+2}$ and the boundary conformal field theory of the world
volume $p$-brane dynamics, have relied on the particular case of
$p=3$ branes of IIB strings on AdS$_5$. This is because the $N=4$
superconformal field theory can be defined in this case for
arbitrary values of the gauge coupling, in view of the exceptional
ultraviolet properties of four-dimensional $N=4$ superconformal
field theory (vanishing beta function).

Unlike this case, the superconformal field theories of M2 and M5
branes, which are dual to M-theory on AdS$_{4(7)}
\times$S$_{7(4)}$, are only understood as strongly coupled
conformal field theories, where the conformal fixed point is only
defined in a formal way. However, the powerful constraints of
superconformal invariance allow one, even in these cases, to
extract some general information, which is supposed to remain
valid in the fully-fledged non-perturbative theory.

The most popular example is the comparison \cite{AOY} of the
spectrum of the 1/2 BPS operators (sometimes called, by an abuse
of language, chiral primary operators) of the superconformal
algebra with the so-called Kaluza-Klein states \cite{GvNW} of
$D=11$ supergravity on AdS$_{4(7)} \times$S$_{7(4)}$ \cite{vNAll}.
The 1/2 BPS operators are the simplest short UIRs of
superconformal algebras, since they correspond to superfields with
maximal shortening (1/2 of the $\theta$s missing). In the case of
the superalgebra $\mbox{OSp}(8^{*}/4)$, which is the subject of
the present paper, 1/2 BPS operators in superspace have been
considered in \cite{Hproc,FS2}. These UIRs have a simple
description \cite{GvNW,GT} in terms of the oscillator method
developed in the 1980s in the pioneering papers of Ref.
\cite{GunAll}. However, many more short UIRs exist for generic
superconformal algebras, even in interacting field theories (such
as $N=4$ super-Yang-Mills (SYM) theory in $D=4$) and these have
been systematically classified by using superfields of different
kinds \cite{FS1,HHowe}. In particular, BPS multiplets are
described by Grassmann (G-)analytic superfields, a generalization
of the familiar notion of chiral superfields of $N=1, \, D=4$
superconformal algebra $\mbox{SU}(2,2/1)$.

For conformal field theories in $D=6$, non-perturbative information on
their superconformal regime can be extracted by superconformal OPEs,
which encode many of the non-perturbative definitions of a generic
superconformal field theory.

Actually, such an approach is a revival of the so-called ``bootstrap program''
of the 1970s, when conformal techniques were popular in connection to the study
of the short-distance behaviour of scale-invariant field theories (for a review
see \cite{Pisa}).

The main new fact, in the case where conformal symmetry is merged with
supersymmetry, are the so-called ``non-renormalization theorems'' of
superconformal field theories, which have a simple explanation in terms of the
existence, in these theories, of a wide class of operators with protected
``conformal dimension'' due to shortening conditions \cite{FF,dp,Minw2} of
the Harish-Chandra modules \cite{HCh,FeFo} of the UIRs.

In the present paper we focus the analysis on the M5 $(2,0)$
conformal theory, based on the $\mbox{OSp}(8^{*}/4)$
superconformal algebra, and we analyse the OPE of two
superconformal 1/2 BPS primary operators. Following \cite{AES,ES},
this is done by resolving the UIR spectrum of operators, which
have a non-vanishing three-point function with the two 1/2 BPS
operators. As we explain later on, superconformal symmetry
uniquely fixes such three-point functions, and further implies
selection rules for the third operator. Partial results on $D=6$
$(2,0)$ three-point functions appear in the literature \cite{P6},
also in the case of 1/2 BPS operators \cite{DHref38}.

{}From standard properties of superconformal field theories this
technique allows one to analyse $n$-point correlator functions by
multiple OPEs and, in some cases, to extract remarkable
non-perturbative information, which has a counterpart on the
supergravity side, such as the extremal, next-to-extremal and
near-extremal correlators
\cite{DHoFrMaMaRa,DHoFrMaMaRa1,D'Hoker:2000dm,DP}.

Many analogies and differences with the four-dimensional case
emerge from this analysis. In $D=4$ the superconformal algebra
$\mbox{PSU}(2,2/4)$ admits three series of UIRs \cite{dp}, one
series (A) with a continuous spectrum of the conformal dimension
and two isolated series (B and C) with fixed (``quantized")
dimension. On the other hand, the UIRs of the $D=6$ superconformal
algebra $\mbox{OSp}(8^{*}/4)$ fall into four distinct series
\cite{Minw2}, one continuous series (A) and three isolated series
(B, C and D) \cite{FS2,FS1}.

Let us denote by
\begin{equation}\label{calD}
  {\cal
D}(\ell;J_1,J_2,J_3;a_1,a_2)
\end{equation}
the quantum numbers of a generic supermodule of $\mbox{OSp}(8^{*}/4)$. Here
$\ell$ is the conformal dimension, $J_1,J_2,J_3$ are the Dynkin labels of the
$D=6$ Lorentz group $\mbox{SU}^{*}(4)\equiv  Spin\;\mbox{SO}(1,5)$ and
$a_1,a_2$ are the Dynkin labels of the R symmetry group $\mbox{USp}(4)$ $\equiv
Spin\; \mbox{SO}(5)$. The four unitary series correspond to:
\begin{equation}\label{6.33'}
  \begin{array}{lll}
   A) & J_1,J_2,J_3\mbox{ unrestricted,} \qquad  & l \geq 6 + \frac{1}{2}
(J_1+2 J_2+3 J_3) + 2(a_1+a_2)
 \\
 B) & J_3 = 0, \qquad & l = 4 + \frac{1}{2} (J_1+2 J_2)
+ 2(a_1+a_2)  \\
 C) & J_2=J_3=0, \qquad & l = 2 + \frac{1}{2} J_1 + 2(a_1+a_2)  \\
 D) & J_1=J_2=J_3=0,
\qquad & l = 2 (a_1+a_2)
  \end{array}
\end{equation}
As we see, the  three isolated series B, C, D occur for $J_1 J_2
J_3 = 0$, while the  continuous series A exists for all values of
$J_1,J_2,J_3$. Operators from  series A saturating the unitarity
bound, as well as all the operators in the isolated series B and C
correspond to ``semishort" superfields with some missing powers of
$\theta$s in their expansion \cite{FS1}. The isolated series D
contains 1/2 and 1/4 BPS states realized in terms of G-analytic
superfields independent of two or one $\theta$, respectively. The
1/2 BPS states correspond to $a_1=0$, i.e., to the symmetric
traceless representations of $\mbox{SO}(5)$. Massless conformal
fields (``supersingletons" \cite{ff2}) belong both to series D
with $a_1+a_2=1$ and to series C with $a_1=a_2=0$
\cite{GT,Fernando:2001ak}.

The main result of the present paper consists in selection rules
for the three-point function of two 1/2 BPS operators ${\cal
D}(2m;0,0,0;0,m)$ and ${\cal D}(2n;0,0,0;0,n)$ with a third
operator\\ ${\cal D}(\ell;0,s,0;a_1,a_2)$. \footnote{Note that the
$\mbox{SU}^{*}(4)$ representations $[0,s,0]$ or, equivalently, the
symmetric traceless rank $s$ tensors of $\mbox{SO}(1,5)$, are the
only Lorentz irreps allowed to appear in the OPE of two scalar
operators.} What we find is summarized below.

The allowed values of $a_1,a_2$ are
$$
a_1=2j\,,\; \; a_2=m+n-2k-2j \quad\mbox{with}\quad 0 \leq k \leq
\min(m,n), \quad 0 \leq j \leq \min(m,n)-k     \,.
$$
Depending on the value of $k$, there are three distinct cases:
$$
  \begin{array}{llll}
   {\it (i)} & k=0 & s=0, \; \ell=2(m+n) &
             \mbox{;\ series D, BPS states, 1/2 BPS if $j=0$} \\
    {\it (ii)} & k=1 & s=0, \; \ell=2(m+n-2) & \mbox{;\ series D, BPS states, 1/2 BPS if
$j=0$} \\
     &  & s \geq 0, \; \ell = 4 + s +
2(m+n-2) & \mbox{;\ series B with $J_1=0$, semishort multiplets} \\
   {\it (iii)} & k \geq 2 & &\mbox{;\ no restrictions on the UIR  (continuous } \\
   & & & \mbox{\quad spectrum possible)}
  \end{array}
$$
The important fact is that in cases {\it (i)} and {\it (ii)} the operators
have ``protected dimension'' because they are either BPS or
``semishort representations''.

A similar phenomenon has recently been observed in the case of
$N=4$, $D=4$ SYM theory \cite{ES}. However, there the semishort
operators are at the threshold of the unitarity bound of the
continuous series rather than at an isolated point. Operators at
the threshold of continuous series or at isolated points can be
realized as products of ``supersingletons" \cite{FS1}. The
surprising fact is that for $D=6$ the  bilinear supersingleton
composite operators belong to the isolated series ({\it protected
dimension}) rather than to the continuous series ({\it unprotected
dimension}) as is the case in $D=4$. Remarkably, the continuous
unitary series starts at the three-singleton threshold and this is
one of the mysteries of $D=6$ superconformal field theory.

The above selection rules have some dramatic consequences for
extremal $n$-point correlators of 1/2 BPS states (i.e. those for
which $a^k_1=0$, $k=1,\ldots,n$ and $a_2^1 = \sum_{k=2}^n a_2^k$).
By multiple superconformal OPE we show that extremal correlators
correspond to the exchange of only 1/2 BPS states, which confirms
the non-renormalization conjecture of Ref. \cite{DP}. One may say
that the extremal correlators of 1/2 BPS operators correspond to a
sub-field theory solely built in terms of 1/2 BPS states.
Incidentally, we remark that BPS operators form a ``ring'' under
multiplication. A possible r\^ole of the algebra of BPS states was
also put forward some time ago in conjunction with other
aspects of string theory and M-theory \cite{MH}.

This paper is organized as follows: \\ In Section 2 we recall $D=6$ (2,0)
superconformal fields, harmonic superspace and short representations
constructed in terms of the 1/2 BPS supersingleton (the tensor multiplet). We
also explain how semishort superfields are constructed in terms of
supersingletons. In Section 3 we study superconformal three-point functions of
two 1/2 BPS operators and a third, {\it a priori} general operator. The crucial
property of such three-point functions is that they are uniquely determined by
conformal supersymmetry. Imposing the shortness conditions at two points, we
derive selection rules for the operator at the third point. In this way we
establish the OPE spectrum of 1/2 BPS operators. In Section 4 we apply these
results to $n$-point extremal and to four-point next-to-extremal correlators
of 1/2 BPS operators and compare with the AdS supergravity non-renormalization
predictions.

\section{Representations of $\mbox{OSp}(8^{*}/4)$ and $D=6$ $(2,0)$ \\
harmonic superspace}

A simple method of constructing the series of UIRs of
superconformal algebras was proposed in \cite{FS1}. The idea is to
start with the massless supermultiplets (``supersingletons").
Then, by multiplying them in all possible ways, we are able to
construct the four series of UIRs of $\mbox{OSp}(8^{*}/4)$ found
in \cite{Minw2}. An essential ingredient in this construction is
harmonic superspace \cite{GIK1}.\footnote{The harmonic superspace
formulation of the basic $D=6$ $(2,0)$ 1/2 BPS multiplet, the
tensor multiplet, and of composite 1/2 BPS operators made out of
it, was first proposed in \cite{Hproc} (see also \cite{Htensor}).}
In this section we give a brief summary, paying special attention
to the UIRs that have fixed conformal dimension, namely the BPS
and the semishort multiplets. They will play an important r\^ole
in our OPE analysis in Section 3.

\subsection{Supersingletons of  $\mbox{OSp}(8^{*}/4)$}

There exist three types of massless multiplets in six dimensions corresponding
to ultrashort UIRs (supersingletons) of $\mbox{\mbox{OSp}}(8^*/4)$ (see, e.g.,
\cite{GT}). All of them can be formulated in terms of constrained superfields
as follows \cite{FS2}.

{\sl (i)} The first type comes in two species. The first one is described by a
superfield $W^{i}(x,\theta)$, $i=1,\ldots,4$ in the fundamental irrep $[1,0]$
of the R symmetry group $\mbox{USp}(4)$. It satisfies the on-shell constraint
\begin{equation}\label{6.19}
  D^{(k}_\alpha W^{i)}=0 \qquad \Rightarrow \ {\cal
D}(2;0,0,0;1,0) \, ,
\end{equation}
which reduces the superfield to the $\mbox{\mbox{OSp}}(8^*/4)$ UIR indicated in
(\ref{6.19}). The spinor covariant derivatives $D^i_\alpha$ obey the
supersymmetry algebra
\begin{equation}\label{6.21}
  \{ D^i_\alpha, D^j_\beta\} =
-2i\Omega^{ij}\gamma^\mu_{\alpha\beta}\partial_\mu\;.
\end{equation}
It is convenient to make the non-standard choice of the symplectic matrix
$\Omega^{ij}=-\Omega^{ji}$ with non-vanishing entries $\Omega^{14}=\Omega^{23}=
- 1$. The chiral $\mbox{SU}^*(4)$ spinors satisfy a
pseudo-reality condition of the type $\overline{D_\alpha^i} =
\Omega^{ij}D_j^\beta c_{\beta\alpha} \, ,$ where $c$ is a $4\times 4$ unitary
``charge conjugation" matrix.

The second species, the so-called  $(2,0)$ tensor multiplet
\cite{HST,bsvp}, is of special interest in $D=6$ SCFT, since it is
related to the basic degrees of freedom of the M5 brane
world-volume theory \cite{AGMOO}. It is described by an
antisymmetric traceless real superfield
$W^{ij}=-W^{ji}=\overline{W_{ij}}$, $\Omega_{ij}W^{ij}=0$
($\mbox{USp}(4)$ irrep $[0,1]$) subject to the on-shell constraint
\begin{equation}\label{6.190}
  D^{(k}_\alpha W^{i)j}=0 \qquad \Rightarrow \ {\cal
D}(2;0,0,0;0,1) \,.
\end{equation}

{\sl (ii)} The second type of supersingletons is described by a (real) Lorentz
scalar and $\mbox{USp}(4)$ singlet superfield, $w(x,\theta)$ obeying the
constraint
\begin{equation}\label{6.23}
 D^{(i}_{[\alpha} D^{j)}_{\beta]} w = 0 \qquad \Rightarrow \ {\cal
D}(2;0,0,0;0,0)\;.
\end{equation}

{\sl (iii)} Finally, there exists an infinite series of multiplets described by
$\mbox{USp}(4)$ singlet superfields with $n$ totally symmetrized external
Lorentz spinor indices, $w_{(\alpha_1\ldots\alpha_n)}(x,\theta)$ (they can be
made real in the case of even $n$) obeying the constraint
\begin{equation}\label{6.24}
  D^i_{[\beta}w_{(\alpha_1]\ldots\alpha_n)} = 0 \qquad \Rightarrow \ {\cal
D}(2+n/2;n,0,0;0,\ldots,0)\;.
\end{equation}

The constraints (\ref{6.19}), (\ref{6.190}), (\ref{6.23}) and (\ref{6.24})
restrict the $\theta$ expansion of the above superfields to just a few massless
fields. In this sense the supersingletons are ``ultrashort" superfields.

\subsection{Harmonic superspace, Grassmann analyticity and BPS multiplets}

The massless multiplets {\sl (i)} admit an alternative formulation
in harmonic superspace. The advantage of this formulation is that
the constraints (\ref{6.19}), (\ref{6.190}) become conditions for
Grassmann analyticity which simply mean that the superfield does
not depend on one or more of the Grassmann variables
$\theta^\alpha$.

We introduce harmonic variables describing the coset
${\mbox{USp}}(4)/\mbox{U}(1)\times \mbox{U}(1)$:
\begin{equation}\label{6.25}
  u\in {\mbox{USp}}(4): \qquad u^I_iu^i_J = \delta^I_J\;,
\ \ u^I_i \Omega^{ij}u^J_j = \Omega^{IJ}\;, \ \  u^I_i= (u^i_I)^*\;.
\end{equation}
Here the indices $i,j$ belong to the fundamental representation of
${\mbox{USp}}(4)$ and $I,J$ are labels corresponding to the $\mbox{U}(1)\times
\mbox{U}(1)$ projections. The harmonic derivatives
\begin{equation}\label{6.26}
  D^{IJ} = \Omega^{K(I}u^{J)}_i \frac{\partial}{\partial u^K_i}
\end{equation}
are compatible with the definition (\ref{6.25}) form the algebra
of ${\mbox{USp}}(4)$ realized on the indices $I,J$ of the
harmonics.

Let us now project the defining constraint (\ref{6.19}) with the harmonics
$u^1_k u^1_{i}$:
\begin{equation}\label{6.27}
D^1_\alpha W^{1} = 0 \, ,
\end{equation}
where $D^{1}_\alpha = D^k_\alpha u^{1}_k$ and
$W^{1}=W^{i}u^1_{i}$. Indeed, the constraint (\ref{6.19}) now
takes the form of a {\it G-analyticity
condition}.\footnote{Grassmann analyticity \cite{GIO} is a
generalization of the concept of chiral superfields familiar from
$N=1, \, D=4$ supersymmetry. In six dimensions all superfields are
chiral, so G-analyticity remains the only non-trivial notion.} It
is integrable because $\{D^1_\alpha, D^1_\beta\} = 0$, as follows
from (\ref{6.21}). This allows us to find an {\it analytic basis}
in superspace:
\begin{equation}\label{AB}
  x^{\alpha\beta}_A  = x^{\alpha\beta} - i\left(\theta^{1[\alpha} \theta^{4\beta]}
  +  \theta^{2[\alpha} \theta^{3\beta]} \right)\,, \qquad \theta^{I\alpha} =
  \theta^{i\alpha}u^I_i
\end{equation}
in which
\begin{equation}\label{6.210}
  D^1_\alpha =  \frac{\partial}{\partial\theta^{4\alpha}}\,.
\end{equation}
Then the solution to (\ref{6.27}) is a {\it Grassmann analytic} superfield,
which does not depend on one of the four odd (spinor) coordinates,
$\theta^{4\alpha}$ (hence the name {\it 1/4 BPS short superfield}):
\begin{equation}\label{6.28}
W^{1}(x_A,\theta^1,\theta^2,\theta^{3},u)\,,
\end{equation}
but it depends on the harmonic variables. Note that the {\it analytic
superspace} with coordinates $x_A,\theta^1,\theta^2,\theta^{3},u$  is closed
under the Poincar\'e ($Q$) supersymmetry transformations,
\begin{equation}\label{Qsusy}
  \delta_Q  x^{\alpha\beta}_A  =  -2i(\theta^{1[\alpha}
  \epsilon^{4\beta]} + \theta^{2[\alpha}
  \epsilon^{3\beta]}  + \theta^{3[\alpha}
  \epsilon^{2\beta]})\,, \qquad  \delta_Q  \theta^{I\alpha} = \epsilon^{I\alpha}
  \equiv  \epsilon^{i\alpha}u^I_i\,, \quad I=1,2,3     \,,
\end{equation}
as well as under the conformal ($S$) supersymmetry transformations.

The G-analytic superfield (\ref{6.28}) is infinitely reducible under
$\mbox{USp}(4)$ because of its harmonic dependence. In order to make it
irreducible (or harmonic short), we impose the harmonic differential conditions
\begin{equation}\label{6.29}
   D^{11}W^{1} = D^{12}W^{1} = D^{13}W^{1} = D^{22}W^{1} = 0\,.
\end{equation}
The harmonic derivatives (\ref{6.26}) appearing in (\ref{6.29})
correspond to the raising operators (positive roots) of
$\mbox{USp}(4)$, thus (\ref{6.29}) can be interpreted as the
defining condition of the HWS of the irrep $[1,0]$. Alternatively,
if a complex parametrization of the harmonic coset
${\mbox{USp}}(4)/\mbox{U}(1)\times \mbox{U}(1)$ is used, eqs.
(\ref{6.29}) take the form of harmonic (H-)analyticity
(Cauchy-Riemann) conditions.

The tensor multiplet (\ref{6.190}) can be reformulated in a similar way.
Projecting the defining constraint with the harmonics $u^K_k u^1_{i}u^2_j$,
$K=1,2$, we obtain:
\begin{equation}\label{6.270}
D^1_\alpha W^{12} = D^2_\alpha W^{12}= 0
\end{equation}
whose solution, in an appropriate modification of the analytic basis
(\ref{AB}), is a harmonic superfield that does not depend on half of the
Grassmann variables:
\begin{equation}\label{6.280}
W^{12}(x_A,\theta^1,\theta^2,u) \, .
\end{equation}
It is subject to the same $\mbox{USp}(4)$ irreducibility conditions
(\ref{6.29}) as $W^1$ above, but now they define the HWS of the irrep $[0,1]$.

The G-analytic superfields $W^{1}(\theta^1,\theta^2,\theta^{3})$ and
$W^{12}(\theta^1,\theta^2)$ are the simplest examples of the two kinds of {\it
BPS short multiplets} of  $\mbox{OSp}(8^{*}/4)$. The latter are described by
superfields which do not depend on a fraction of the total number of Grassmann
variables. Thus, $W^1$ is 1/4 BPS short and $W^{12}$ is 1/2 BPS short.

Further BPS short superfields can be obtained by multiplying the above two
types of supersingletons and imposing the $\mbox{USp}(4)$ irreducibility
conditions (\ref{6.29}):
\begin{equation}\label{s1}
  W^{[a_1,a_2]} = [W^{1}(\theta^1,\theta^2,\theta^{3})]^{a_1}
   [W^{12}(\theta^1,\theta^2)]^{a_2}  \qquad \Rightarrow \ {\cal
D}(2(a_1+a_2);0,0,0;a_1,a_2)\;.
\end{equation}
Note that the powers of each type of supersingleton correspond to the Dynkin
labels of the $\mbox{USp}(4)$ irrep. This can be seen by looking at the leading
component ($\theta=0$) of the superfield. The differential conditions
(\ref{6.29}) reduce it to a harmonic monomial:
\begin{equation}\label{s101}
  W^{[a_1,a_2]}(\theta=0) = C^{i_1\cdots i_{a_1+a_2}\; j_1\cdots j_{a_2}}
  \; u^1_{i_1}\cdots u^1_{i_{a_1+a_2}} u^2_{j_1} \cdots u^2_{j_{a_2}} \, ,
\end{equation}
where the indices of the coefficient tensor $C$ form the (anti)symmetrized
traceless combinations of the Young tableau $(a_1+a_2,a_2)$. In general,
$W^{[a_1,a_2]}$ is 1/4 BPS short unless $a_1=0$ when it becomes 1/2 BPS short.
In the next subsection we show that (\ref{s1}) realizes the most general series
of BPS short multiplets of $\mbox{OSp}(8^{*}/4)$. The case of 1/2 BPS operators
is of particular importance for us, and we therefore introduce the following
simplified notation for them:
\begin{equation}\label{12bps}
  {\cal W}^m \equiv  W^{[0,m]} = [W^{12}(\theta^1,\theta^2)]^{m}  \qquad \Rightarrow \ {\cal
D}(2m;0,0,0;0,m)\;.
\end{equation}

Concluding this subsection we point out that there exists an alternative way of
constructing a subclass of 1/4 BPS short multiplets that is relevant to our
discussion of OPEs of 1/2 BPS operators in Section 3. To obtain it we first
reformulate the tensor multiplet in an equivalent form by projecting the
defining constraint (\ref{6.190}) with the harmonics $u^K_k u^1_{i}u^3_j$,
$K=1,3$:
\begin{equation}\label{6.2700}
D^1_\alpha W^{13} = D^3_\alpha W^{13}= 0\quad \Rightarrow \quad  W^{13} =
W^{13}(\theta^1,\theta^3)\,.
\end{equation}
Thus we obtain a new 1/2 BPS superfield depending on a different half of the
odd variables. However, this time it is not a HWS of a $\mbox{USp}(4)$ irrep
since the raising operator $D^{22}$ does not annihilate it:
\begin{equation}\label{s2}
  D^{22} W^{13}= W^{12} \,.
\end{equation}
Further, the product of superfields $W^{12}W^{13}$ has exactly the same content
as $[W^1]^2$ because they depend on the same $\theta$s and at $\theta=0$ one
finds $C^{ijkl}u^1_iu^1_{[j}u^2_ku^3_{l]} = C^{ij}u^1_iu^1_j$. In general,
\begin{equation}\label{s3}
  W^{[2j,p]} = [W^{12}(\theta^1,\theta^2)]^{p+j}
   [W^{13}(\theta^1,\theta^3)]^{j}  \qquad \Rightarrow \ {\cal
D}(2(2j+p);0,0,0;2j,p)\;.
\end{equation}
Once again, this is a 1/4 BPS superfield unless $j=0$ when it becomes 1/2 BPS.
The only difference from the general BPS series (\ref{s1}) is that in
(\ref{s3}) the first $\mbox{USp}(4)$ Dynkin label is even.

\subsection{Series of UIRs of ${\mbox{OSp}}(8^*/4)$ and semishort multiplets}

A study of the most general UIRs of ${\mbox{OSp}}(8^*/4)$ (similar to the one
of Ref. \cite{dp} for the case of ${\mbox{SU}}(2,2/N)$) is presented in Ref.
\cite{Minw2}. We can construct these UIRs by multiplying the three types of
supersingletons above:
\begin{equation}\label{6.32}
  w_{\alpha_1\ldots\alpha_{p_1}}w'_{\beta_1\ldots\beta_{p_2}}
w''_{\gamma_1\ldots\gamma_{p_3}}\; w^k\; W^{[a_1,a_2]}\,.
\end{equation}
Here $p_1\geq p_2 \geq p_3$ and we have used three different copies of the
supersingleton with spin, so that the spinor indices can be arranged to form an
${\mbox{SU}}^*(4)$ UIR with Young tableau $(p_1,p_2,p_3)$ or Dynkin labels
$J_1=p_1-p_2,J_2=p_2-p_3,J_3=p_3$. We thus obtain the four distinct series in
(\ref{6.33'}).\footnote{Comparing the UIRs of ${\mbox{OSp}}(8^*/4)$ to those
of the $N=4$ superconformal algebra in four dimensions $\mbox{PSU}(2,2/4)$
\cite{dp}, we remark that both include one continuous series and that the
number
of discrete series corresponds to the rank of the Lorentz group.} The reason
why the conformal dimension in series A is continuous, while it is
``quantized'' in the other three
series, has to do with the unitarity of the corresponding
irreps.\footnote{The author of \cite{Minw2} conjectures the existence of a
``window" of continuous dimensions $2 +\frac{1}{2}J_1+2(a_1+a_2)\leq \ell \leq 4
+\frac{1}{2}J_1+2(a_1+a_2)$ if $J_2=J_3=0$, but this has not been proved.}

Series D contains the 1/4 or 1/2 BPS short multiplets (obtained by dropping all
the $w$ factors in (\ref{6.32})). Series A generically contains ``long"
multiplets, unless the unitarity bound is saturated \cite{Minw2} (this
corresponds to setting $k=0$ in (\ref{6.32})). In series B and C some
``semishortening" takes place. In the realization (\ref{6.32}) this is easily
seen by using the on-shell constraints on the elementary supersingletons. The
full identification of such ``semishort" multiplets is given in \cite{FS1}.
Here we restrict ourselves to the case of series B with $J_1=0$, the only one
relevant to the OPE analysis in Section 3. Let us first set $a_1=a_2=0$, i.e.
no BPS factor appears in (\ref{6.32}). Then we have two possibilities, $J_2\neq
0$ and $J_2=0$.

If $J_2\neq 0$ we take (\ref{6.32}) with only two supersingletons with equal
spin $(p_1=p_2=J_2\neq0)$,
\begin{equation}\label{s4}
  {\cal O}_{\alpha_1\ldots\alpha_{J_2}\; \beta_1\ldots\beta_{J_2}} =
  w_{\alpha_1\ldots\alpha_{J_2}} \; w'_{\beta_1\ldots\beta_{J_2}}
  \ \rightarrow \ \ell = 4 +J_2 \,.
\end{equation}
With the help of the on-shell constraint (\ref{6.24}) we obtain the following
``conservation law"
\begin{equation}\label{6.3312}
  \epsilon^{\delta\gamma\alpha_1\beta_1} D^i_\gamma\;
  {\cal O}_{\alpha_1\ldots\alpha_{J_2}\; \beta_1\ldots\beta_{J_2}} = 0\,.
\end{equation}

If $J_2=0$ we keep only two scalar supersingletons in (\ref{6.32}),
\begin{equation}\label{s5}
  {\cal O} = w^2 \ \rightarrow \ \ell = 4 \,.
\end{equation}
Using the on-shell constraint (\ref{6.23}) we obtain the following
``conservation law"
\begin{equation}\label{6.3316}
  \epsilon^{\delta\gamma\beta\alpha}D^{(i}_\gamma D^j_\beta
D^{k)}_\alpha\; {\cal O} = 0\;.
\end{equation}

In both cases above, the semishort superfield is a bilinear in the
supersingletons, just like the conserved current $J_\mu(x) = \bar\phi(x)
\partial_\mu\phi(x) - \phi(x) \partial_\mu \bar\phi(x)$, $\partial^\mu
J_\mu=0$, constructed out of two massless scalar fields. Indeed, the
conservation laws (\ref{6.3312}) and (\ref{6.3316}) imply that some of the
components of the semishort superfield are conserved (spin-)tensors. However,
we can relax these conservation conditions by assigning ${\mbox{USp}}(4)$
quantum numbers to the above ``currents". This is done by multiplying them by a
BPS short superfield (\ref{s1}) (or (\ref{s3})):
\begin{equation}\label{s6}
  {\cal O}^{[a_1,a_2]}_{\alpha_1\ldots\alpha_{J_2}\; \beta_1\ldots\beta_{J_2}} =
  w_{\alpha_1\ldots\alpha_{J_2}} \; w'_{\beta_1\ldots\beta_{J_2}}\; W^{[a_1,a_2]}
  \ \rightarrow \ \ell = 4 +J_2 + 2(a_1+a_2) \,,
\end{equation}
\begin{equation}\label{s7}
  {\cal O}^{[a_1,a_2]} = w^2\; W^{[a_1,a_2]} \ \rightarrow \ \ell = 4 + 2(a_1+a_2) \,.
\end{equation}
These new semishort superfields satisfy the corresponding ${\mbox{USp}}(4)$
projections of the conservation laws (\ref{6.3312}) and (\ref{6.3316}), for
example
\begin{equation}\label{s8}
  \epsilon^{\delta\gamma\alpha_1\beta_1} D^1_\gamma\;
  {\cal O}^{[a_1,a_2]}_{\alpha_1\ldots\alpha_{J_2}\;
  \beta_1\ldots\beta_{J_2}} =
  0\,,
\end{equation}
\begin{equation}\label{s9}
  \epsilon^{\delta\gamma\beta\alpha}D^{1}_\gamma D^1_\beta
D^{1}_\alpha\; {\cal O}^{[a_1,a_2]} = 0
\end{equation}
in the case of a 1/4 BPS factor in (\ref{s6}) and (\ref{s7}).

The new weaker constraints do not imply the presence of conserved
(spin-)tensors among the components of the semishort superfields,
they just eliminate some of these components. In other words, some
powers of $\theta$s are missing in the expansion, but not entire
$\theta^\alpha$s as in the case of a G-analytic superfield, hence
the distinction between BPS short and semishort multiplets.

A very important point is that the semishort superfields, like the
BPS ones, have fixed quantized dimension. However, unlike the BPS
superfields, the ``conservation" conditions on the semishort ones
may be broken by dynamical effects in an interacting field theory,
so their quantized dimension can in principle be affected by
renormalization.\footnote{An interesting discussion of this point
is given in the recent paper \cite{HH} in the context of $N=4$ SYM
in four dimensions.} At the same time we should stress that there
is a ``dimension gap" between the semishort multiplets from series
B and the continuous series A. Therefore it seems impossible to
change the status of these semishort operators from ``protected"
to ``unprotected" by small radiative corrections. In this sense
the six-dimensional case is quite different from the
four-dimensional.

\section{Three-point functions involving two 1/2 BPS operators}

In this section we investigate the OPE of two 1/2 BPS operators (\ref{12bps}),
${\cal W}^m\; {\cal W}^n$. To this end we construct all possible superconformal
three-point functions:
\begin{equation}\label{5.1'}
  \langle {\cal W}^m(x,\theta,1) \,
  {\cal W}^n(y,\zeta,2) \,
  {\cal O}^{{\cal D}}(z,\lambda,3) \rangle \, ,
\end{equation}
where ${\cal O}^{{\cal D}}$ is an {\it a priori} arbitrary
operator carrying an $\mbox{OSp}(8^{*},4)$ UIR labeled by the
quantum numbers ${\cal D}$ from eq. (\ref{calD}).  The three sets
of space-time, Grassmann and harmonic variables are denoted by
$x,y,z$; $\theta,\zeta,\lambda$; $1^I_i,2^I_i,3^I_i$,
respectively. It is understood that the appropriate G-analytic
basis (cf. (\ref{AB})) is used at points 1 and 2.

Conformal supersymmetry uniquely fixes such three-point functions. Indeed, the
superfunction (\ref{5.1'}) depends on half of the Grassmann variables at points
1 and 2 and on a full set of such variables at point 3. Thus, the total number
of odd variables exactly matches the number of supersymmetries (Poincar\'e $Q$
plus special conformal $S$). Therefore there exist no nilpotent superconformal
invariants and the complete $\theta,\zeta,\lambda$ expansion of (\ref{5.1'}) is
determined from the leading ($\theta=\zeta=\lambda=0$) component. The latter is
the three-point function of two scalars and one tensor field, and is fixed by
conformal invariance.

Apart from G-analyticity, the  three-point function (\ref{5.1'}) should also
satisfy the requirement of $\mbox{USp}(4)$ irreducibility (H-analyticity) at
points 1 and 2. This leads to selection rules for the third UIR for the
following reason. The coefficients in, for instance, the $\theta$ expansion
at point 1
depend on the harmonics in a way that matches the harmonic $\mbox{U}(1) \times
\mbox{U}(1)$ charges of $\theta^{1,2}_\alpha$. The crucial point is that some
of these coefficients may be harmonic singular, thus violating the requirement
of H-analyticity. Demanding that such singularities be absent excludes some
UIRs at point 3.

\subsection{Two-point functions}

Before discussing the three-point function (\ref{5.1'}) itself, consider first
the two-point function of two tensor multiplets:
\begin{equation}\label{5.101}
  \langle W^{12}(x,\theta^{1,2},1)
  W^{12}(y,\zeta^{1,2},2) \rangle \, .
\end{equation}
It has the leading component
\begin{equation}\label{5.3}
  \langle W^{12} W^{12}\rangle_{\theta=\zeta=0} = \frac{(1^{12}2^{12})}{(x-y)^4}
\end{equation}
where, for example
\begin{equation}\label{5.4}
  (1^{12}2^{12}) \equiv \epsilon^{ijkl} 1^1_i 1^2_j 2^1_k 2^2_l
\end{equation}
is the only possible $\mbox{USp}(4)$ invariant combination of the harmonics at
points 1 and 2 carrying the required $\mbox{U}(1) \times \mbox{U}(1)$ charges.

The irreducibility (H-analyticity) conditions at point 1 are easily checked:
$\partial^{22}_1$ trivially annihilates the numerator and $\partial^{13}_1$
does so because of the antisymmetrization; similarly at point 2. The space-time
dependence in (\ref{5.3}) follows from the fact that $W$ has dimension 2 and
spin 0.

The G-analytic superfunction (\ref{5.101}) depends on $2+2$ spinor coordinates,
as many as the $Q$ supersymmetry parameters. So, it is sufficient to use only
$Q$ supersymmetry to restore the odd variable dependence starting from
(\ref{5.3}). Also, we will restrict our attention to harmonic singularities at
point 1 (point 2 is similar), whence it is sufficient to restore the $\theta$
dependence only. Thus, we stay in a coordinate frame in which $\zeta=0$. Since
this condition does not involve the harmonics at point 1, it cannot introduce
singularities with respect to them. In such a frame there is a residual $Q$
supersymmetry given by the condition
\begin{equation}
  \delta'_Q \zeta^{1,2} = 0 \Rightarrow {\epsilon'}^i =
  (2_3^i 2^3_j + 2_4^i 2^4_j) \epsilon^j
   \label{5.102} \, .
\end{equation}
In deriving (\ref{5.102}) we used the definition (\ref{6.25}) of
the harmonics, from which it follows that $u_4 = -u^1$, $u_3 = - u^2$
(the raising and lowering of the $I$ and $i$ indices is
independent).

Now, in the analytic basis, $x$ transforms as follows:
\begin{equation}\label{5.103}
  \delta_Q x^{\alpha \beta} = -2i(\theta^{1[\alpha}
  \epsilon^{4\beta]} + \theta^{2[\alpha}
  \epsilon^{3\beta]})
\end{equation}
(compare to (\ref{AB}) and (\ref{Qsusy})). Replacing the
parameters in (\ref{5.103}) by the residual ones from
(\ref{5.102}) we can find $\delta'_Q x$. Then the combination
\begin{equation}\label{5.104}
  x^{\alpha \beta} +
  i (a_{11} \, \theta^{1[\alpha}
  \theta^{1 \beta]} - 2 \, a_{12} \, \theta^{1[\alpha}
  \theta^{2 \beta]} + a_{22} \, \theta^{2[\alpha}
  \theta^{2 \beta]} )
\end{equation}
with
\begin{equation}
a_{11} = \frac{(1^{24} 2^{12})}{(1^{12}2^{12})}, \; \; \; a_{12} =
\frac{(1^{14} 2^{12})}{(1^{12}2^{12})} = - \frac{(1^{23}
2^{12})}{(1^{12}2^{12})}, \; \; \; a_{22} = - \frac{(1^{13}
2^{12})}{(1^{12}2^{12})}
\end{equation}
is invariant under the residual $Q$ supersymmetry. Noting that
$\delta'_Q y=0$, we can write the two-point function (\ref{5.101})
in the frame $\zeta=0$ as a coordinate shift of its leading
component (\ref{5.3}):
\begin{equation} \label{5.105}
  \langle W^{12}(\theta)
  W^{12}(\zeta=0) \rangle = \exp\left\{ - \frac{i}{4} \, \left( a_{11}
  (\theta^1 \partial_x \theta^1) - 2 a_{12} (\theta^1 \partial_x \theta^2) +
a_{22} (\theta^2 \partial_x \theta^2) \right) \right\}\,
\frac{(1^{12}2^{12})}{(x-y)^4} \,.
\end{equation}

The coefficients $a_{11}, a_{12}, a_{22}$ in (\ref{5.105}) introduce the
harmonic singularity
\begin{equation} \label{5.1055}
U = \frac{1}{(1^{12}2^{12})}
\end{equation}
since $(1^{12}2^{12})=0$ when points 1 and 2 coincide. The identity
\begin{equation} \label{5.106}
a_{11} \, a_{22} - a_{12}^2 = \frac{(1^{34} 2^{12})}{(1^{12}2^{12})}
\end{equation}
may be used to simplify the expansion of the exponential. While the product of,
say, $a_{11}$ and $a_{22}$ contains two such denominators, the r.h.s. of
(\ref{5.106}) has only one power of the singularity. We find in this way that
no higher singularity than $U^2$ occurs in the exponential, and that all terms
with $U^2$ contain at least one operator $\square_x$. Since $\square (x^2)^{-3}
\sim \delta^6(x)$, the harmonic-singular terms in the expansion of
(\ref{5.105}) are space-time contact terms. We conclude that the two-point
function is regular as long as $x\neq y$, owing to its harmonic numerator. This
will not automatically be so for the three-point functions.

In the following we will also need the two-point function
\begin{equation} \label{5.107}
\langle W^{12}(1) W^{13}(2) \rangle
\end{equation}
for the two alternative realizations (\ref{6.280}) and (\ref{6.2700}) of the
tensor multiplet. In the frame $\zeta=0$ this is obtained from (\ref{5.105}) by
replacing the harmonic $2^2$ by $2^3$ everywhere.

\subsection{Three-point functions $\langle {\cal W}^{m}(1) \,
  {\cal W}^{n}(2) \, {\cal O}^{\cal D}(3) \rangle$}

In close analogy with the two-point functions above, here we
investigate the three-point functions (\ref{5.1'}) starting with
their leading component, then restoring the dependence on $\theta$
and finally imposing H-analyticity at point 1.

The $\mbox{USp}(4)$ irrep carried by ${\cal O}^{\cal D}(3)$ should be in the
decomposition of the tensor product of the two irreps at points 1 and 2:
\begin{equation}\label{5.2}
[0,m] \otimes [0,n] \, = \, \bigoplus_{k=0}^n \, \bigoplus_{j=0}^{n-k} \, \,
[2j, m+n - 2k -2j] \, ,
\end{equation}
where we have assumed that $m\geq n$. The first Dynkin label being even, the
irrep $[2j,p]$ can be realized as a product of $W^{12}$s and  $W^{13}$s,
recall (\ref{s3}). This suggests to construct the $\mbox{USp}(4)$ structure of
the function (\ref{5.1'}) as a product of two-point functions of $W$s.

Apart from the $\mbox{USp}(4)$ quantum numbers the operator ${\cal
O}(3)$ also carries spin and dimension. Since the leading
components at points 1 and 2 are scalars, the Lorentz irrep at
point 3 must be a symmetric traceless tensor of rank $s$ or,
equivalently, an $\mbox{SU}^*(4)$ irrep $[0,s,0]$. The
corresponding conformal tensor structure is built out of the
vector
\begin{equation}\label{5.8}
  Y^\mu = \frac{(x-z)^\mu}{(x-z)^2} - \frac{(y-z)^\mu}{(y-z)^2} \,.
\end{equation}
All in all, the leading term is:
\begin{eqnarray}
 &&\hskip-1.5cm \langle {\cal W}^m(1) \, \nonumber
  {\cal W}^n(2) \,
  {\cal O}^{(\ell; \, 0, s, 0; \, 2j, m+n-2k-2j)}(3)\rangle_{0} = \\
 &&\quad \left[\frac{(1^{12}2^{12})}{(x-y)^4}\right]^k
  \left[\frac{(1^{12}3^{12})}{(x-z)^4}\right]^{m-j-k}
   \left[\frac{(2^{12}3^{12})}{(y-z)^4}\right]^{n-j-k} \nonumber\\
  &&\quad \times \left\{\left[\frac{(1^{12}3^{12})}{(x-z)^4}\right]
   \left[\frac{(2^{12}3^{13})}{(y-z)^4}\right]
  - \left[\frac{(1^{12}3^{13})}{(x-z)^4}\right]
   \left[\frac{(2^{12}3^{12})}{(y-z)^4}\right]\right\}^j  \nonumber\\
  &&\quad \times\ (Y^2)^{\frac{\ell-s}{2}-m-n+2k}\ Y^{\{\mu_1}\cdots
  Y^{\mu_s\}} \, ,  \label{5.9}
\end{eqnarray}
where $\{\mu_1\cdots\mu_s\}$ denotes traceless symmetrization. The $3^2,3^3$
antisymmetrization in the factor $\{ \ldots \}^j$ reflects the properties of
the $\mbox{USp}(4)$ Young tableau $(m+n-2k,2j)$ or, equivalently, the harmonic
irreducibility constraints at point 3.

To study the harmonic singularities at point 1 we restore the
dependence on $\theta$, keeping $\zeta = \lambda = 0$. We need
both $Q$ and $S$ supersymmetry to reach this new frame. The
harmonics $1^I_i$ do not participate in the frame fixing, so that
there is no danger of creating harmonic singularities at point 1.
Next, under conformal boosts the vector $Y^\mu$ (\ref{5.8})
transforms homogeneously with parameters involving only $z$. We
need to find a superextension of $Y^\mu$ with the same properties:
It should be invariant under $Q+S$ supersymmetry at points 1 and 2
(and covariant at point 3, but we do not see this in the present
frame). Remarkably, the combination (\ref{5.104}) that was $Q$
invariant in the two-point case turns out to be $Q+S$ invariant in
the new frame. Thus, performing the shift (\ref{5.104}) of the
variable $x$ in the vectors $Y$ in (\ref{5.9}) we obtain the
desired superextension. In addition, the two-point factor $(1^{12}
2^{12})/(x-y)^4$ in (\ref{5.9}) undergoes the same shift. The
factors involving $(x-z)^2$ are supersymmetrized by a similar
shift, which, as explained in the preceding subsection, does not
induce a harmonic singularity when $1 \rightarrow 3$, at least up
to space-time contact terms.

The harmonic factor $\{ \ldots \}^j$ in (\ref{5.9}) vanishes for $1 \rightarrow
2$, but cannot compensate the singularities of the type $U$ (\ref{5.1055})
coming from the exponential shift (\ref{5.105}). To show this, we must
identify the four complex coordinates on the harmonic coset
${\mbox{USp}}(4)/\mbox{U}(1)\times \mbox{U}(1)$ in terms of ${\mbox{USp}}(4)$
invariant combinations of the harmonics. Then it becomes clear that the
singularity in $U$ and the ``zero" in the factor $\{ \ldots \}^j$ correspond to
different directions on the coset. We do not present the details here.

So, we can concentrate on the exponential shift (\ref{5.105}). This involves
the terms
\begin{eqnarray} \label{5.10}
\exp\left\{ - \frac{i}{4} \, \left( a_{11}
  (\theta^1 \partial_x \theta^1) - 2 a_{12} (\theta^1 \partial_x \theta^2) +
a_{22} (\theta^2 \partial_x \theta^2) \right) \right\} \, \times \\
\left[ \frac{(1^{12}2^{12})}{(x-y)^4} \right]^k
(Y^2)^{\frac{\ell-s}{2}-m-n+2k}\ Y^{\{\mu_1}\cdots
  Y^{\mu_s\}} \, . \nonumber
\end{eqnarray}
The factor $(1^{12}2^{12})^k$ can suppress singularities. But
here, as opposed to the two-point case, the presence of the $Y$
terms will not always allow this. We distinguish three cases:

{\it (i)}  If $k=0$ a singularity occurs already in the $\theta \theta$ term.
In order to remove it we require:
\begin{equation}\label{5.14}
  \partial^\nu_x \left\{(Y^2)^{\frac{\ell-s}{2}-m-n}\ Y^{\{\mu_1}\cdots
  Y^{\mu_s\}} \right\} = 0 \, ,
\end{equation}
which implies
\begin{equation}\label{5.15}
  s=0\,, \quad \ell = 2(m+n) = 2(a_1+a_2) \, ,
\end{equation}
where $[a_1,a_2]$ is the $\mbox{USp}(4)$ irrep at point 3. This constraint
sends the expression (\ref{5.10}) to unity, reducing the three-point function
to a product of two-point functions:
\begin{eqnarray}
  && \langle {\cal W}^m(1) \,
  {\cal W}^n(2) \,
  {\cal O}^{(2(m+n); \, 0, 0, 0; \, 2j, m+n-2j)}(3)\rangle = \label{5.1400} \\
  &&\qquad \langle W^{12}(1) W^{12}(3) \rangle^{m-j} \
  \langle W^{12}(2) W^{12}(3) \rangle^{n-j} \nonumber\\
  &&\qquad \times \left\{\langle W^{12}(1) W^{12}(3) \rangle
  \ \langle W^{12}(2) W^{13}(3) \rangle -
  \langle W^{12}(1) W^{13}(3) \rangle
  \ \langle W^{12}(2) W^{12}(3) \rangle
    \right\}^j  \,. \nonumber
\end{eqnarray}
The operator at the third point is seen to belong to series D from
(\ref{6.33'}), i.e. it is 1/4 BPS short if $j\neq 0$ or 1/2 BPS short if
$j=0$.

{\it (ii)} If $k=1$ the singularity is in the $U^2$ terms, all of which involve
at least one operator $\square_x$. In order to remove it we demand:
\begin{equation}\label{5.11}
  \square_x \left\{(x-y)^{-4}
  \ (Y^2)^{\frac{\ell-s}{2}-m-n+2}\ Y^{\{\mu_1}\cdots
  Y^{\mu_s\}} \right\} = 0\,.
\end{equation}
This equation is identically satisfied in two cases: \\
{\it (ii.a)} We can have $\ell = -s + 2(m+n-2) = -J_2 + 2(a_1+
a_2)$. Looking at (\ref{6.33'}) we see that this is only
compatible with series D, whence $s=J_2=0$. So, the first solution
is
\begin{equation} \label{firstk1}
s=0, \qquad \ell = 2(a_1+a_2) \, ,
\end{equation}
which again corresponds to a BPS short operator at point 3.

{\it (ii.b)}  We may put $\ell=s + 2(m+n-2) + 4$, i.e.
\begin{equation} \label{secondk1}
\ell = 4 +  J_2 + 2(a_1+a_2) \, ,
\end{equation}
and we recognize the B series from (\ref{6.33'}) with $J_1=0$. It is
therefore expected that the operator at point 3 is ``semishort",
i.e. that it satisfies the constraints (\ref{s8}) when $s\neq0$
or (\ref{s9}) when $s=0$. Indeed, in the next subsection we shall
prove this.

{\it (iii)} If $k\geq 2$, the expression (\ref{5.10}) is
completely regular, so we obtain no selection rules from harmonic
analyticity.

\subsection{Semishortening at the third point}

In case {\it (ii.b)}, the operators ${\cal O}(3)$ belong to series B and should
thus obey the semishortness constraints (\ref{s8}) or (\ref{s9}). If we use
the analytic basis (\ref{AB}) at point 3, the spinor derivative $D^1_\alpha$
becomes a partial derivative, see (\ref{6.210}). Then the semishortness
conditions on the three-point function constructed above take the following
form:
\begin{eqnarray}
 & s \neq 0: \qquad
& \epsilon^{\delta \gamma \alpha_1 \beta_1} \frac{\partial}{\partial\lambda^{4
\gamma}} \,\langle {\cal W}^m(1) {\cal W}^n(2) {\cal O}_{\alpha_1 \ldots
\alpha_s \ \beta_1 \ldots \beta_s}
(z_A,\lambda,3)\rangle = 0 \, ,  \label{const1}  \\
& s=0: \qquad & \epsilon^{\delta \gamma \alpha \beta}
\frac{\partial}{\partial\lambda^{4 \gamma}}\frac{\partial}{\partial\lambda^{4
\alpha}}\frac{\partial}{\partial\lambda^{4 \beta}}  \ \langle {\cal W}^m(1) {\cal
W}^n(2) {\cal O} (z_A,\lambda,3)\rangle = 0 \, . \label{const2}
\end{eqnarray}
We will now verify that these conditions are indeed satisfied.

As we explained earlier, the complete $\theta,\zeta,\lambda$ dependence of the
three-point function can be restored starting from the leading component
(\ref{5.9}). The factors $[\cdots]^{\cdots}$ can easily be upgraded to
two-point functions of the type (\ref{5.101}) and (\ref{5.107}). From
(\ref{s3}) we know that any product of $W^{12}(3)$ and $W^{13}(3)$ is
annihilated by $\partial_{4 \alpha}$, hence it trivially satisfies
(\ref{const1}), (\ref{const2}). Thus, we only need to impose these conditions
on the supersymmetrization of the factor $(Y^2)^{\frac{\ell-s}{2}-m-n+2}\
Y^{\{\mu_1}\cdots Y^{\mu_s\}}$ (recall that $k=1$ in case  {\it (ii.b)}). There
exists a standard method \cite{ParkOsb,P6} for constructing the supercovariant
version of $Y$, but the resulting expressions are rather complicated.
Fortunately, we are only interested in the dependence of the $Y$ factor on
$\lambda^{4\alpha}$ at point 3 which is very easy to reconstruct.

Using $Q$ and $S$ supersymmetry, translations and conformal boosts we can
choose a frame where only the coordinates $x^{\alpha\beta}$,
$\lambda^{4\alpha}$ and the three sets of harmonics remain:
\begin{equation} \label{frame}
y \rightarrow \infty,\qquad z = 0, \qquad \theta^{1,2} = 0, \qquad \zeta^{1,2}
= 0, \qquad \lambda^{1,2,3} = 0\,.
\end{equation}
This frame is harmonic singular, but now we are not interested in
harmonic analyticity at point 3. The residual transformation
preserving the frame is
\begin{equation}
{\delta'}_{Q+S} x^{\alpha \beta} = 0\,, \qquad {\delta'}_{Q+S} \lambda^{4
\alpha} = (x^{\alpha \beta} + 2i A_{44} \lambda^{4 \alpha} \lambda^{4 \beta})
\xi^4_\beta  \,, \qquad {\delta'}_{Q+S} (1^I_i,2^I_i,3^I_i) =0 \, .
\end{equation}
Here $\xi^4_\beta$ is the transformation parameter and
\begin{equation}
A_{44} = \frac{(3^1 2_{[3})(1^{[1} 2_{4]})(1^{2]}
3_4)}{(1^{12}2^{12})} \, ,
\end{equation}
where, for instance, $(3^1 2_{3})\equiv 3^1_i 2^i_3\,$, etc.

It is then clear that the combination
\begin{equation} \label{QShat}
\hat x^{\alpha \beta} \, = \, x^{\alpha \beta} + 2i A_{44}
\lambda^{4 \alpha} \lambda^{4 \beta}
\end{equation}
has a homogeneous transformation law,
\begin{equation}\label{trs}
{\delta'}_{Q+S} \, \hat x^{\alpha \beta} =  \Lambda\, \hat x^{\alpha \beta} +
\Sigma^{[\alpha}{}_\gamma \, \hat x ^{\beta] \gamma}\,, \qquad \Lambda = -i
A_{44} \lambda^{4 \alpha} \xi^4_\alpha\,, \quad \Sigma^{\alpha}{}_\gamma = 4i
A_{44} \lambda^{4 \alpha} \xi^4_\gamma - \mbox{trace}\,.
\end{equation}
Here $\Lambda$ and $\Sigma$ are the coordinate-dependent parameters of a
dilatation and of a Lorentz transformation, correspondingly.

Next, in the frame (\ref{frame}) the vector $Y$ (\ref{5.8}) becomes $Y^\mu =
x^\mu/x^2$. Replacing $x$ by $\hat x$ from (\ref{QShat}) in $Y$, we obtain a
vector with a covariant transformation law. Thus, the $\lambda^{4\alpha}$
dependence of the $Y$ factor in (\ref{5.9}) is restored  by a simple coordinate
shift, the result of which is
\begin{eqnarray}
  &&\hskip-1cm (Y^2)^{\frac{\ell-s}{2}-m-n+2}\ Y^{\{\mu_1}\cdots Y^{\mu_s\}} \ \rightarrow  \nonumber\\
  && \left\{1 - \frac{i}{2} A_{44} \lambda^{4} \partial_x \lambda^{4} +A^2_{44}
(\lambda^{4})^4 \, \square_x \right\}\ (x^2)^{-\frac{\ell+s}{2}+m+n-2}\
x^{\{\mu_1}\cdots x^{\mu_s\}}\,. \label{shiftexp}
\end{eqnarray}

Now we can easily impose conditions (\ref{const1}), (\ref{const2}). The first
one amounts to the requirement that the rank $s+1$ tensor ($s\neq0$)
\begin{equation}\label{cst1}
  \partial_x^{\nu} \left[ (x^2)^{-\frac{\ell+s}{2}+m+n-2}\
x^{\{\mu_1}\cdots x^{\mu_s\}} \right]
\end{equation}
is symmetric and traceless. It is verified when $\ell = s + 2(m+n)$, just as
expected, recall (\ref{secondk1}) (note that  the $(\lambda^{4})^4$ term in
(\ref{shiftexp}) automatically vanishes in this case). The second condition
only concerns the $(\lambda^{4})^4$ term in (\ref{shiftexp}):
\begin{equation}\label{cst2}
  \square_x (x^2)^{-\frac{\ell}{2}+m+n-2}  = 0
\end{equation}
which holds when $\ell = 2(m+n)-2$ (case ({\it ii.a})) or when $\ell = 2(m+n)$
(case ({\it ii.b})).

We conclude that when the operator at point 3 has the right quantum numbers to
belong to series B, the three-point function automatically satisfies the
corresponding semishortening condition. This is reminiscent of the situation in
bosonic $D$-dimensional CFT. There the three-point function of two scalars of
equal dimension and a vector of dimension $\ell$, $\langle\phi(1)\phi(2)
j^\mu(3)\rangle$ is automatically conserved, $\langle\phi(1)\phi(2)
j^\mu(3)\rangle \overleftarrow{\partial}_{3\mu} =0$, when $\ell$ takes the
appropriate value $\ell=D-1$.

\section{Extremal and next-to-extremal correlators}

In this section we discuss certain classes of $n$-point
correlation functions of 1/2 BPS operators ${\cal W}^m \equiv
[W^{12}]^m$:
\begin{equation}\label{e1}
  \langle {\cal W}^{m_1}(1) {\cal W}^{m_2}(2) \cdots {\cal W}^{m_n}(n) \rangle \, .
\end{equation}
According to the terminology introduced in \cite{DHoFrMaMaRa} they are called
\begin{eqnarray}
 \mbox{``extremal" if } && m_1 =\sum_{i=2}^n m_i \nonumber\\
 \mbox{``next-to-extremal" if } && m_1 =\sum_{i=2}^n m_i - 2 \label{e2} \\
 \mbox{``near-extremal" if } && m_1 =\sum_{i=2}^n m_i - 2k\,,
 \quad k\geq 2\,.       \nonumber
\end{eqnarray}
Using AdS supergravity arguments, it was conjectured in \cite{DP} that the
extremal and next-to-extremal correlators are not renormalized and factorize
into products of two-point functions, whereas the near-extremal ones are
renormalized but still factorize into correlators with fewer numbers of points.
With the help of the OPE results from Section 3, we prove here the
non-renormalization and factorization of the $n$-point extremal correlators as
well as the non-renormalization of the next-to-extremal four-point correlator.
We also speculate about a possible way to understand the factorization of
near-extremal correlators.

\subsection{The extremal case}

We begin by the simplest case, which is a four-point extremal correlator.
It can be represented as the convolution of two OPEs:
\begin{eqnarray}
  && \langle {\cal W}^{m_1}(1) {\cal W}^{m_2}(2) {\cal W}^{m_3}(3) {\cal W}^{m_4}(4) \rangle =
  \label{e3} \\
  &&\qquad \sum \hskip-13pt\int_{5,5'}
  \langle {\cal W}^{m_1}(1) {\cal W}^{m_2}(2) {\cal O}(5) \rangle \
  \langle  {\cal O}(5)  {\cal O}(5') \rangle^{-1} \
\langle  {\cal O}(5') {\cal W}^{m_3}(3) {\cal W}^{m_4}(4)\rangle \, , \nonumber
\end{eqnarray}
where the sum goes over all possible operators that appear in the intersection
of the two OPEs. Owing to the orthogonality of different operators the inverse
two-point function $\langle  {\cal O}(5)  {\cal O}(5') \rangle^{-1}$ only
exists if ${\cal O}(5)$ and ${\cal O}(5')$ are identical.\footnote{In CFT every
operator ${\cal O}$ has the so-called ``shadow" operator $\tilde {\cal O}$
such that the two can form a non-diagonal two-point function of the type
$\langle {\cal O}(1) \tilde {\cal O}(2) \rangle = \delta(1-2)$. However, these
``shadows" only exceptionally have physical dimension (i.e. do not violate the
unitarity bound), so they usually need not be considered in an OPE. It is easy
to show that this is the case here.} To find out their spectrum, we first
examine the $\mbox{USp}(4)$ quantum numbers. From (\ref{5.2}) we see that
\begin{eqnarray}
 {\cal W}^{m_1}(1) {\cal W}^{m_2}(2) \ \rightarrow \ {\cal O}(5)\,: &&
 \bigoplus_{k=0}^{m_2}
  \bigoplus_{j=0}^{m_2-k}[2j,m_1+m_2-2j-2k]\, , \nonumber\\
  {\cal W}^{m_3}(3) {\cal W}^{m_4}(4) \ \rightarrow \ {\cal O}(5')\,: &&
 \bigoplus_{k'=0}^{m_4}
  \bigoplus_{j'=0}^{m_4-k'}[2j',m_3+m_4-2j'-2k']\,, \label{e4}
\end{eqnarray}
where we have assumed $m_3 \geq m_4$. Since in the extremal case
$m_1=m_2+m_3+m_4$ (recall (\ref{e2})), the intersection is given by the
following conditions:
\begin{equation}\label{e5}
  j=j'\,, \qquad 0 \leq k' = k-m_2 \leq 0\,,
\end{equation}
whose only solution is
\begin{equation}\label{e6}
  k=m_2 \ \Rightarrow \ j=j'=0\,, \qquad k'=0 \,.
\end{equation}
Further, we deduce from (\ref{5.1400}) that $k'=0$ and $j'=0$ imply that ${\cal
O}(5')$, and by orthogonality, ${\cal O}(5)$ must be identical 1/2 BPS
operators,
\begin{equation}\label{e7}
  {\cal O} =  {\cal W}^{m_3+m_4}\,.
\end{equation}
Finally, in this particular case the three-point functions in (\ref{e3})
degenerate into products of two two-point functions (recall (\ref{5.1400})), so
(\ref{e3}) becomes
\begin{eqnarray}
  && \hskip-2cm \langle {\cal W}^{m_1}(1) {\cal W}^{m_2}(2) {\cal W}^{m_3}(3) {\cal W}^{m_4}(4) \rangle
  \label{e8} \\
  &=&\int_{5'}  \langle W(1) W(2) \rangle^{m_2} \ \int_{5} \langle W(1) W(5) \rangle^{m_3+m_4} \
  \langle W(5) W(5') \rangle^{-(m_3+m_4)}    \nonumber\\
  &&\qquad \times  \langle W(5') W(3) \rangle^{m_3}\
   \langle W(5') W(4) \rangle^{m_4}      \nonumber\\
  &=& \langle W(1) W(2) \rangle^{m_2} \ \langle W(1) W(3) \rangle^{m_3} \
  \langle W(1) W(4) \rangle^{m_4} \,. \nonumber
\end{eqnarray}
This clearly shows that the extremal four-point correlator factorizes into a
product of two-point functions. In other words, it always takes its free (Born
approximation) form, so it stays {\it non-renormalized}.

The generalization of the above result to an arbitrary number of points is
straightforward. We explain it on the example of a five-point extremal
correlator. This time we have to perform three consecutive OPEs, so the analog
of (\ref{e3}) is
\begin{eqnarray}
  && \langle {\cal W}^{m_1}(1) \cdots {\cal W}^{m_5}(5) \rangle =
  \label{e9} \\
  &&\qquad
  \langle {\cal W}^{m_1}(1) {\cal W}^{m_2}(2) {\cal O}(6) \rangle \bullet
  \langle  {\cal O}(6) {\cal W}^{m_3}(3) {\cal O}(7) \rangle \bullet
\langle  {\cal O}(7)  {\cal W}^{m_4}(4) {\cal W}^{m_5}(5)\rangle \nonumber \, ,
\end{eqnarray}
where $\bullet$ denotes the convolution with the inverse two-point functions at
the internal points 6 and 7. As before, we start by examining the
$\mbox{USp}(4)$ quantum numbers. The sum of the Dynkin labels of an irrep is a
$\mbox{U}(1)$ charge. In the tensor product of two irreps the charge of the
product ranges from the sum to the difference of the two charges, e.g. in
$[0,m_1]\otimes [0,m_2]$ we obtain values between $m_1+m_2$ and
$m_1-m_2=m_3+m_4+m_5$. Moving along the chain (\ref{e9}) from left to right,
and each time choosing the minimal value, when we arrive at the last pair of
points, we are just able to match the maximal value $m_4+m_5$ coming from the
tensor product of the last two irreps. Thus the only possible chain of irreps
is as follows:
\begin{equation}\label{e10}
  [0,m_1]\otimes [0,m_2] \ \rightarrow \
  [0,m_3+m_4+m_5]\otimes [0,m_3] \ \rightarrow \
  [0,m_4+m_5] \ \leftarrow \  [0,m_4]\otimes [0,m_5]\,.
\end{equation}
Note that at each step the first Dynkin label is 0.

Now, let us start moving from right to left. Using (\ref{5.2}) we see that the
first step corresponds to $k=j=0$, so it produces a single 1/2 BPS operator in
the OPE, ${\cal O}(7) = {\cal W}^{m_4+m_5}(7)$. Consequently, at the second
step we again have the OPE of two 1/2 BPS operators producing yet another 1/2
BPS operator ${\cal O}(6) = {\cal W}^{m_3+m_4+m_5}(6)$. If $n>5$ this process
goes on until we reach the first pair of points. We conclude that the
$n$-point extremal correlators of 1/2 BPS operators are based on exchanges of
$(n-2)$ 1/2 BPS operators only. We may call this a ``field theory of 1/2 BPS
operators".

Finally, just as in the four-point case (\ref{e8}) above, the three-point
functions in (\ref{e9}) become degenerate (products of two two-point functions)
and we achieve the expected factorization of the extremal correlator:
\begin{eqnarray}
  && \hskip-2cm \langle {\cal W}^{m_1}(1) \cdots {\cal W}^{m_5}(5) \rangle
  \label{e11} \\
  &=&\int_{6',7'}  \langle W(1) W(2) \rangle^{m_2} \
  \int_{6} \langle W(1) W(6) \rangle^{m_3+m_4+m_5} \
  \langle W(6) W(6') \rangle^{-(m_3+m_4+m_5)}    \nonumber\\
  &&\qquad \times  \langle W(6') W(3) \rangle^{m_3} \
  \int_{7} \langle W(6') W(7) \rangle^{m_4+m_5} \
  \langle W(7) W(7') \rangle^{-(m_4+m_5)}    \nonumber\\
  &&\qquad \times  \langle W(7') W(4) \rangle^{m_4}\
   \langle W(7') W(5) \rangle^{m_5}      \nonumber\\
  &=& \langle W(1) W(2) \rangle^{m_2} \ \langle W(1) W(3) \rangle^{m_3} \
  \langle W(1) W(4) \rangle^{m_4}\ \langle W(1) W(5) \rangle^{m_5} \,. \nonumber
\end{eqnarray}

\subsection{The next-to-extremal and near-extremal cases}

The situation is considerably more complicated in the next-to-extremal case,
even with just four points, $m_1=m_2+m_3+m_4-2$. Repeating the steps leading to
(\ref{e5}), this time we find the conditions
\begin{equation}\label{e12}
  j=j'\,, \qquad 0 \leq k'-1 = k-m_2 \leq 0 \, ,
\end{equation}
which admit two solutions:
\begin{equation}\label{e13}
   k=m_2-1\,, \quad k'=0\,, \quad \left\{
  \begin{array}{lll}
    j=0 & \rightarrow & {\cal O}^{[0,m_3+m_4]}(5')\quad \mbox{is 1/2 BPS} \\
    j=1 & \rightarrow & {\cal O}^{[2,m_3+m_4-2]}(5')\quad \mbox{is 1/4 BPS}
  \end{array}\right.
\end{equation}
or
\begin{equation}\label{e14}
   k=m_2\,, \quad k'=1\,, \quad j=0 \  \rightarrow \
  {\cal O}^{[0,m_3+m_4-2]}(5')\quad \mbox{is 1/2 BPS or semishort}  \ .
\end{equation}

We see that unlike the extremal case, where only a finite number of 1/2 BPS
operators are exchanged, here one encounters 1/4 BPS and semishort ones. The
latter form an infinite series, so the OPE content is much richer. Still, there
is an important restriction: all the operators in the OPE have protected
integer dimension.\footnote{In fact, this case resembles the conformal partial
wave expansion (or double OPE) of the {\it free} four-point function of
physical scalars of canonical dimension \cite{Pisa}. There one finds an
infinite spectrum of operators of integer dimension (conserved tensors).} To
put it differently, we have shown that no operators of anomalous dimension can
occur in the expansion of the next-to-extremal four-point correlator. Since
renormalization in ultraviolet-finite CFT is associated with the appearance of
anomalous dimensions, we can conclude that the correlator is non-renormalized.

However, showing that the amplitude factorizes into a product of two-point
functions is not so easy now. The reason is that the three-point functions
$\langle {\cal W} {\cal W} {\cal O} \rangle$ themselves no longer factorize,
so evaluating expression (\ref{e3}) implies doing conformal
four-star integrals \cite{Symanzik}. Yet, the calculation may turn out to be
rather trivial, once again because we are only dealing with integer dimensions.
Indeed, in the case at hand the three-point functions $\langle {\cal W} {\cal
W} {\cal O} \rangle$ involve singular distributions of the type $1/(x^2)^k$,
$k\geq3$ with delta-function type singularities. After properly regularizing
the integrals, this is expected to result in the factorization of the
amplitude.

Finally, we could try to apply our arguments to the near-extremal correlators.
In this case tensoring the $\mbox{USp}(4)$ irreps at, for example,
points 1 and 2, and
then going along the chain leaves room for irreps $[2j,m_1+m_2-2k-2j]$ with
$k\geq 2$. In other words, operators with unprotected dimension are allowed to
appear, so the correlator can be renormalized. One might speculate that, for
instance, the near-extremal six-point condition $m_1 =\sum_{i=2}^6 m_i - 4$
will restrict the occurrence of a $k=2$ exchange (and hence of anomalous
dimensions) to only one of the OPEs, the rest still involving operators of
protected dimension. As we just explained, the latter are associated with
singular distributions and thus with trivial integrations, whereas the former
will give rise to a non-trivial four-point function. This is a possible
scenario of the factorization conjectured in \cite{D'Hoker:2000dm,DP}, and it
certainly deserves a careful investigation.

We remark that in \cite{DHoFrMaMaRa1} a different approach was used
to prove the non-renormalization of extremal and next-to-extremal correlators
in four-dimensional SCFT. It consists in constructing directly the $n$-point
superconformal invariant in harmonic superspace and then imposing the harmonic
analyticity conditions. In the extremal case this method leads to the
conclusion that the corresponding invariant is unique and coincides with its
free value.\footnote{In \cite{Htensor} it has been suggested to extend this
method to the six-dimensional extremal case.} However, the constraints obtained
in this way for next-to-extremal correlators are weaker and do not allow us to
decide whether they are renormalized or not. In \cite{DHoFrMaMaRa1} additional
dynamical information was used in coming from the insertion the SYM action into
the correlator. In six dimensions there is no known dynamical principle,
therefore this procedure cannot be applied. Consequently, we can say that the
method based on OPE described in this paper is more powerful, at least where
next-to-extremal correlators are concerned.

\section*{Concluding remarks}
The analysis carried out in this paper actually applies to any $D=6$ $(N,0)$
superconformal algebra OSp$(8^{*}/2N)$. We note that, unlike $D=4$, these
algebras have only one kind of 1/2 BPS states in the $[0,\ldots,0,N]$ of the
R symmetry group USp$(2N)$. We expect to find similar selection rules in all of
these cases.

The same method can also be applied to the $D=3$ $N=8$ superconformal field
theories based on the superalgebra OSp$(8/4,\mathbb{R})$.

An extension of our result, which could be relevant to the more detailed
examination of next-to-extremal and near-extremal correlators, is to construct
three-point functions where only at one point there is a 1/2 BPS operator.
In this case superconformal invariance does not uniquely fix the three-point
functions, but one might still hope to find some selection rules.

\section*{Acknowledgements}
The work of S.F. has been supported in part by the European Commission RTN
network HPRN-CT-2000-00131, (Laboratori Nazionali di Frascati, INFN) and by the
D.O.E. grant DE-FG03-91ER40662, Task C.

\end{document}